\documentclass[esd, manuscript]{copernicus}

\begin{document}
\nolinenumbers
\title{Evidence of Amazon rainforest dieback in CMIP6 models}

\Author[1]{Isobel}{Parry}
\Author[1]{Paul}{Ritchie}
\Author[1]{Peter}{Cox}

\affil[1]{College of Engineering, Mathematics and Physical Sciences, University of Exeter, Exeter, UK, EX4 4QE}

\correspondence{Isobel Parry (ip294@exeter.ac.uk)}

\runningtitle{Evidence of Amazon rainforest dieback in CMIP6 models}

\runningauthor{Parry et al.}

\received{}
\pubdiscuss{} 
\revised{}
\accepted{}
\published{}

\firstpage{1}

\maketitle
\begin{abstract}
Amazon forest dieback is seen as a potential tipping point under climate change. These concerns are partly based-on an early coupled climate-carbon cycle simulation, that produced unusually strong drying and warming in Amazonia. In contrast, the 5$^{th}$ generation Earth System Models (CMIP5) produced few examples of Amazon dieback under climate change. Here we examine results from seven 6$^{th}$ generation models (CMIP6) which include vegetation dynamics, and in some cases interactive forest fires. Although these models typically project increases in area-mean forest carbon across Amazonia under CO$_2$-induced climate change, five of the seven models also produce abrupt reductions in vegetation carbon which indicate localised dieback events. The Northern South America region (NSA), which contains most of the rainforest, is especially vulnerable in the models. These dieback events, some of which are mediated by fire, are preceded by an increase in the amplitude of the seasonal cycle in near surface temperature, which is consistent with more extreme dry seasons. Based-on the ensemble mean of the detected dieback events we estimate that 7+/-5 $\%$ of the NSA region will experience abrupt downward shifts in vegetation carbon per $^{o}$C of global warming above 1.5$^{o}$C.
\end{abstract}

\introduction
A ‘tipping point’ commonly refers to small changes to input levels causing a system to abruptly transition to some alternative (often less desirable) stable state \citep{1}. Future tipping points pose a risk to both natural ecosystems and, by extension, human activities, as they produce abrupt system wide changes that are often difficult or even impossible to reverse \citep{4}. The Amazon rainforest is one example in the climate system that is at risk of experiencing a tipping event, with the possibility of abrupt forest dieback in response to rising global temperatures \citep{19}. Amazon dieback has the potential to accelerate global warming through reducing the Amazon's ability to act as a carbon sink, and releasing carbon dioxide that would lead to additional global warming  \citep{30}. Tipping points may play an important role in the future of our changing climate \citep{3,4}, with previous analysis of CMIP5 models suggesting that multiple regional abrupt transitions could occur for global warming levels less than 2 degree Celsius \citep{5}.

There are several factors which could contribute to a decline in vegetation in the Amazon, including a lengthened dry season, increased fire frequency, and reduced precipitation \citep{10}. The number of extreme hot and dry days in the Amazon is predicted to increase with global warming \citep{9} and the length and intensity of the dry season expected to intensify \citep{10}. Further drying in the Amazon is anticipated from the slowdown of the Atlantic Meridional Overturning circulation due to ice melt causing an influx of fresh water into the North Atlantic \citep{11}. Moisture stress resulting from severe droughts in the Amazon is likely to result in a degree of tree mortality \citep{12}. As the rainforest dries it becomes more vulnerable to fire which, coupled with the increased frequency of fires seen in the Amazon over recent years, could lead the rainforest to pass a tipping point and result in vegetation dieback \citep{10,13}. Anthropogenic deforestation  also contributes to this by reducing dry season rainfall and decreasing the resilience of the forest to climate change, potentially leading to permanent forest loss in some regions of the Amazon \citep{14}. Mechanisms which result in the drying of the Amazon rainforest can therefore be considered to be the main cause of vegetation dieback. 

Dieback tipping events are primarily thought of as bifurcation type tipping points \citep{28}, which occur when external climatic factors reduce the resilience  of a state (e.g. forest) and ultimately cause the system to tip into a new contrasting state (e.g. savannah) \citep{2}. For some bifurcation type tipping points there are generic features of a system that can be detected to indicate the approach of a tipping point \citep{2,15}. The most common example  is ‘critical slowing down’, where a system becomes increasingly slow at recovering from small perturbations as negative feedbacks become overwhelmed by positive ones. Critical slowing down can be observed by increases to the autocorrelation and variance in a state variable \citep{2,15}. A recent study which looked at such signals in satellite-retrieved vegetation greenness, reports evidence of reducing resilience of the Amazon rainforest since 2005 \citep{29}. 

However, previous research into projections of Amazonian vegetation dieback has also suggested that generic EWS such as these fail in the Amazon, but that more system specific indicators may be found \citep{16}. For example, the interannual variability of the atmospheric CO$_2$ concentration, as a function of tropical temperature variability, has been shown to be connected to the sensitivity of tropical carbon to climate change \citep{32,35}. This metric shows trends which are also consistent with reducing resilience of tropical forests \citep{34,33}.

These recent studies \citep{29,33} focus on fairly short observational records. In this paper we look instead at projections from the  latest CMIP6 Earth System Models for evidence of Amazon dieback, and identify a precursor which is based-on longer-term temperature records.

\section{Methods}

\subsection{CMIP6 models, experiment runs and data used.} Climate models that incorporate dynamic vegetation and are from the 6th Phase of the Coupled Model Intercomparison Project CMIP6 were utilised in this study. See Table~\ref{Tab:CMIP6_list} for the corresponding seven CMIP6 climate models. For the purpose of this study we wanted to focus on the climatic drivers alone impacting vegetation and therefore make use of the 1pctCO2 runs. Data from the PIControl runs were also used to determine each model's internal variability. Primarily, we use model output data of the vegetation carbon and surface temperature for the seven climate models in the NSA region. The amplitude of the temperature seasonal cycle in this study is defined as the difference between the maximum and minimum monthly mean for each year. All anomalies presented correspond to the yearly mean relative to the mean of the first ten years, aside from temperature anomalies which correspond to the ten year running mean relative to the mean of the first ten years.

\begin{table}[t]
\caption{CMIP6 models used within this study}
\centering
\begin{tabular}{ cc }
 \hline
 \textbf{Model} & \textbf{Institution}\\
 \hline
 EC-Earth3-Veg &  EC-Earth-Consortium \\
 GFDL-ESM4 & NOAA-GFDL  \\
 MPI-ESM1-2-LR & Max-Planck-Institut f\"{u}r Meteorologie  \\
 NorCPM1 & EarthClim  \\
 TaiESM1 & AS-RCEC  \\
 SAM0-UNICON & Seoul National University\\
 UKESM1-0-LL & Met Office Hadley Centre \\
 \hline
\end{tabular}
\label{Tab:CMIP6_list}
\end{table}

\subsection{Abrupt shift detection algorithm.} The algorithm used to detect abrupt shifts is relatively simple by design. Three criteria must be be fulfilled for a grid point to be identified as containing an abrupt shift in the vegetation carbon. Namely, the vegetation carbon must change by at least 2 kgC/m$^2$ over a 15-year period and that this must contribute to at least 25\% of the overall change in vegetation carbon. Finally, to remove detected abrupt shifts that might be due to a model's high internal variability the mean annual rate of change of the abrupt shift must be at least three times larger than the variability of the rates of change in the unforced control run. 

Grid points where abrupt shifts were detected are subsequently sorted based on the direction of the abrupt shift (positive or negative) and the direction of overall trend (positive or negative). This results in four classifications of abrupt shifts, where analysis focused on the dieback abrupt shifts corresponding to the overall trend and direction of the abrupt shift being negative. This type of abrupt shift can be used as an analogy for a tipping event where a region changes equilibrium state from rainforest to savannah. 

\subsection{EWS for Amazon dieback.} Our system specific EWS for Amazon dieback is to observe high sensitivities of the amplitude of the seasonal cycle to global warming. This is defined as the gradient of a linear regression fit to the amplitude of the temperature seasonal cycle against global warming. For a comparison between grid points with abrupt shifts and those that do not, the regression is fitted against the first 73 years of data corresponding to when CO$_2$ has doubled from pre-industrial levels (noting that most abrupt shifts occur after a doubling of CO$_2$). Area weighted histograms of the sensitivities are calculated to compare the distributions for areas with abrupt vegetation dieback (red) and areas with no abrupt shifts (purple). The mean of the seven model histograms was taken to produce the compiled models histogram in Fig.~\ref{Fig_4}h. The percentage risk of an abrupt dieback shift occurring (Fig.~\ref{Fig_4}i) is calculated as the percentage of abrupt dieback grid points to all grid points for the specified sensitivity.

\section{Results}

\begin{figure*}[t]
	\includegraphics[width=1\linewidth]{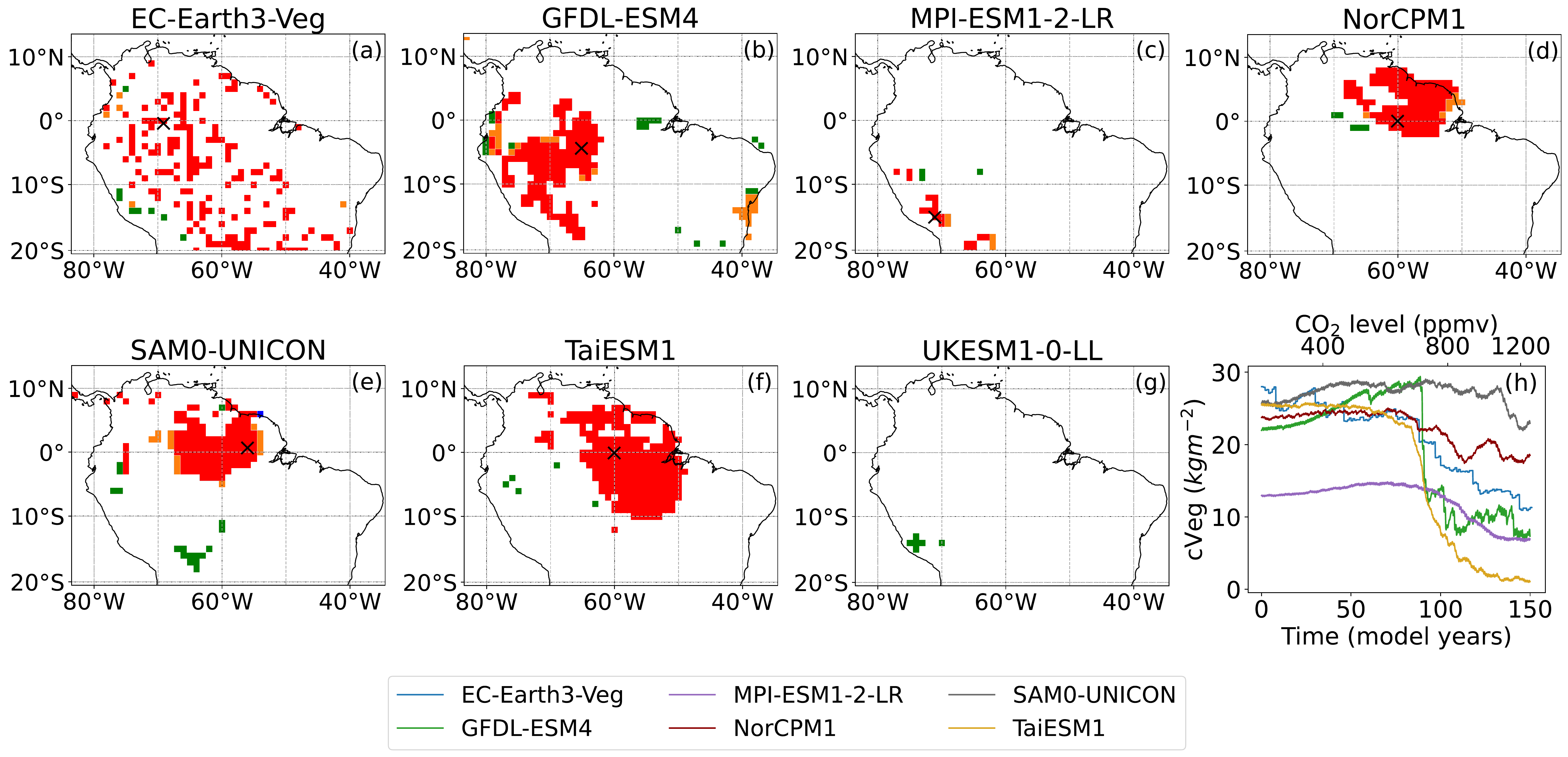}
	\caption{\textbf{Abrupt shifts detected in the Amazon by the described algorithm and example time series for dieback shifts.} \textbf{(a-h)} Maps of detected abrupt shifts. Red grid points indicate detected dieback shifts where the direction of the overall trend (T) and the abrupt shift (AS) are both negative. Green points indicate that the direction of trend and abrupt shift are both positive. Blue points indicate a positive abrupt shift but a negative trend, while orange points indicate a negative abrupt shift associated with a positive trend. \textbf{(h)} Example time series for detected dieback shifts in each model, corresponding to grid points highlighted by black squares.}
	\label{Fig_1}
\end{figure*}

\subsection{Detection of abrupt shifts}
We focus our analysis on detecting Amazon dieback abrupt shifts in seven state-of-the-art climate models, which all enable dynamic vegetation, from the 6th Phase of the Climate Model Intercomparison Project (CMIP6). Specifically, we are interested in climate change induced dieback (rather than direct deforestation) and therefore consider the idealised scenario of $CO_2$ increasing 1\% per year starting from pre-industrial levels. We have defined that for an abrupt shift to be detected a $2kgC/m^2$ change must be observed within a 15-year period and this change must contribute to over a quarter of the change observed in the entire simulation run. We additionally require that the mean annual rate of change is more than 3 times as large as the variability in the rates of change seen in the unforced control runs (see Methods for further details).

The algorithm categorises abrupt shifts into four types dependent on the directions of the abrupt shift (AS) itself and the overall trend (T; change in 5-year means between start and end of the run). Figure~\ref{Fig_1}a--g depicts the types of abrupt shifts detected spatially for the CMIP6 models. Conventional dieback abrupt shifts (red; $T<0$, $AS<0$), which indicate a move towards a savannah state, are predominantly detected. Three models; NorCPM1, TaiESM1 and SAM0-UNICON all show a clustering of dieback abrupt shifts in the north of the Amazon. GFDL-EMS4 also presents a coherent structure with abrupt shifts clustered in central-west Amazonia. Contrast to this, EC-Earth3-Veg shows many abrupt shifts scattered across the Amazon basin. This may be due to the high natural variability that is inherent in this model and which is difficult to distinguish from climate forced abrupt shifts. Interestingly, UKESM1-0-LL displays no dieback events, despite showing large scale dieback in previous CMIP generations \citep{19}. Similarly, very few abrupt shifts are detected in the MPI-ESM1-2-LR model.

Some sample time series of detected dieback abrupt shifts across the models are shown in Fig.~\ref{Fig_1}h. Between the models there is some variation in the general shape of abrupt shift time series, however, most exhibit a change of state from one equilibrium to another that would be expected of a tipping event. 

\begin{figure*}[t]
	\includegraphics[width=1\linewidth]{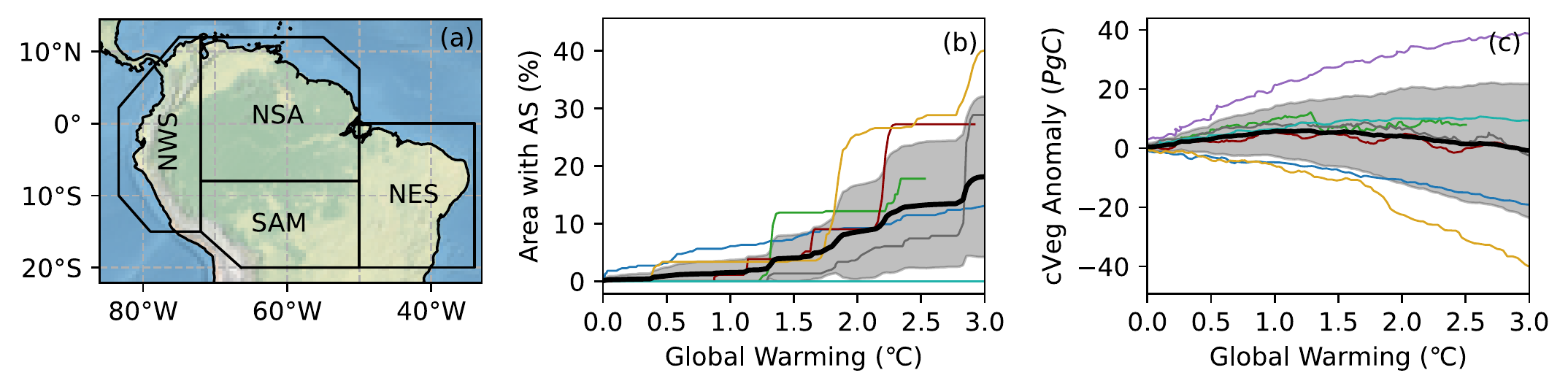}
	\caption{\textbf{Plots showing the evolution of abrupt dieback shifts and vegetation carbon with global warming.} \textbf{(a)} IPCC AR6 defined regions of the Amazon. \textbf{(b)} Plot showing the evolution of the percentage of the NSA region that has experienced a dieback shift with increasing global warming. The plume indicates the error in the averaged compiled models line (black). \textbf{(c)} Plot showing how the vegetation carbon anomaly relative to the mean of the first ten years evolves with global warming. The plume indicates the error in the averaged compiled model line (black).}
	\label{Fig_2}
\end{figure*}

\subsection{Evolution and impact of abrupt shifts in the NSA region}
Henceforth, we restrict our analysis to the IPCC AR6 defined North South America (NSA) region, which contains the majority of the Amazon basin (see Fig.~\ref{Fig_2}a), and features many of the detected abrupt dieback shifts (red points in Fig.~\ref{Fig_1}). Figure~\ref{Fig_2}b shows how the fractional area of the NSA region to experience an abrupt dieback shift evolves for increasing global warming. Some models show clear jumps in the NSA area to experience an abrupt shift reflecting multiple gird points featuring an abrupt shift at a similar level of warming. Strikingly, TaiESM1 shows about 20\% of the NSA region suffering an abrupt dieback event at about $1.7^oC$ warming and by $3^oC$ global warming about 40\% of the NSA region would experience an abrupt shift. The bold black line represents the mean behaviour and the plume variability of all seven models. Although some of the large jumps from individual models can still be identified, the model mean shows a smoother increase in fractional NSA area to undergo an abrupt shift under global warming. There is not a singular temperature threshold, instead the risk of tipping increases (approximately linearly) between $1.3^oC$ and $3^oC$ of warming and reaches approximately 20\% of the NSA region to undergo an abrupt dieback shift. 

Interestingly, when evaluating at the regional scale many of the abrupt shifts do not appear to materialise, despite identifying a significant number of local abrupt shifts, (see Fig.~\ref{Fig_2}c). Only the cluster of abrupt shifts at approximately $1.3^oC$ warming for GFDL-ESM4 appear in the total vegetation carbon anomaly for the NSA region. Furthermore, the CMIP6 models do not even agree on the sign change in vegetation carbon. 

\begin{figure*}[t]
	\includegraphics[width=1\linewidth]{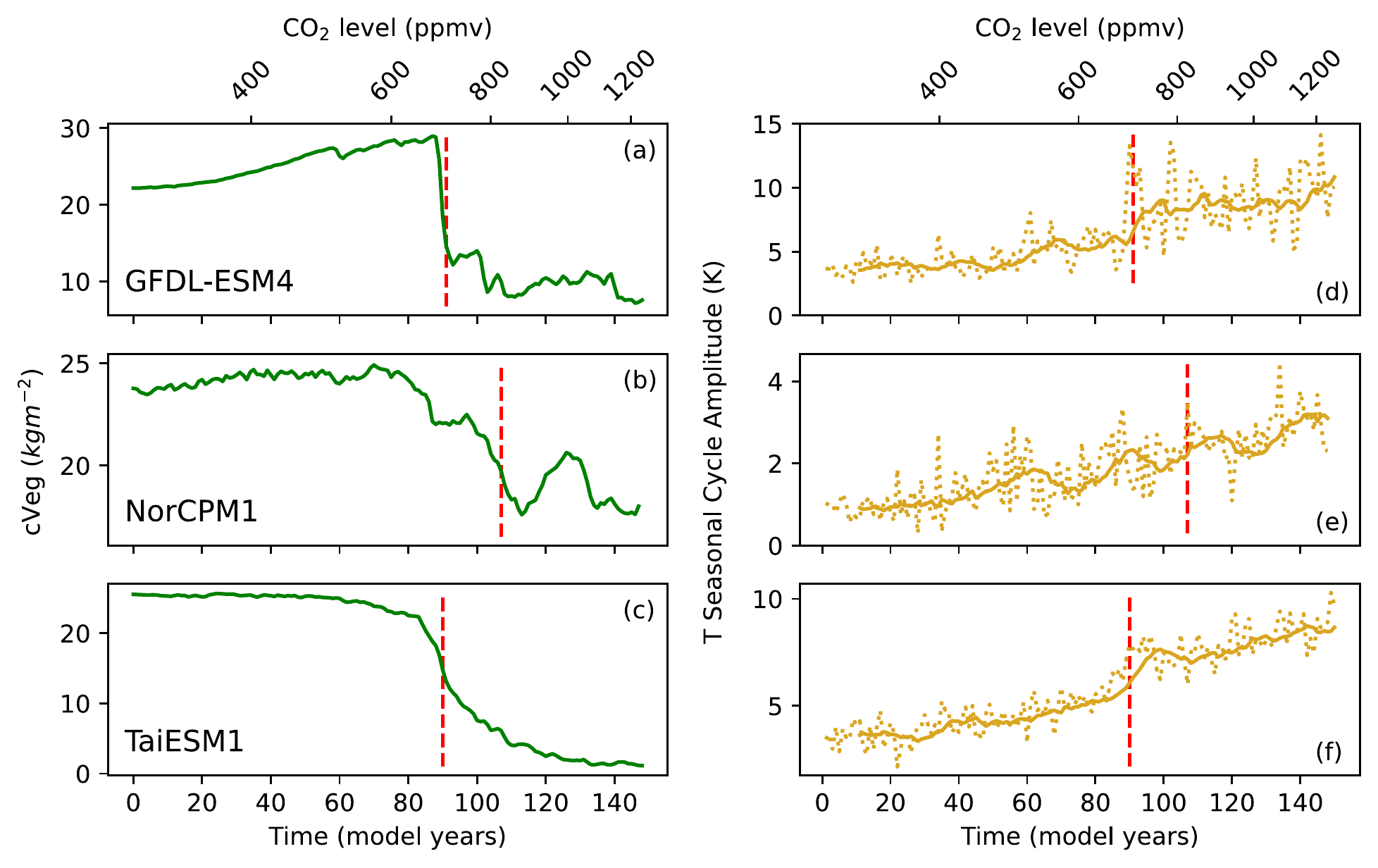}
	\caption{\textbf{Time series showing how the temperature seasonal cycle amplitude evolves over time for grid points which experience abrupt dieback shifts.} \textbf{(a-c)} Time series of selected dieback shifts for three models at grid points 5$^{\circ}$S 65$^{\circ}$W, 0$^{\circ}$ 60$^{\circ}$W and 0$^{\circ}$ 60$^{\circ}$ respectively. The red dotted line indicates the midpoint of the 15 year period where the abrupt shift is detected by the algorithm. \textbf{(d-f)} The change in the amplitude of the temperature seasonal cycle with time and $CO_2$ for these dieback shifts in each model. The solid lines represents the 10 year running average of the seasonal cycle amplitude, while the dotted lines are the yearly data.}
	\label{Fig_3}
\end{figure*}

\subsection{An EWS for Amazonian dieback}
Three identified abrupt shifts, which all occur around a doubling of $CO_2$ in different models, are shown in Fig.~\ref{Fig_3}a--c and all show a change of equilibrium state after the abrupt shift. Initially, vegetation carbon may increase due to the $CO_2$ fertilization effect (c.f. Fig.~\ref{Fig_3}a,b), however there exists a \textit{critical threshold} in the $CO_2$ concentration at which increased temperature and drying overwhelm the positive effect of $CO_2$ and results in an abrupt dieback shift. Fig.~\ref{Fig_3}d--f show the trend in the temperature seasonal cycle associated with these three grid points. An increasing trend is observed in the amplitude of the seasonal cycle in the lead up to an abrupt shift for each point, suggesting that this increase in variability may be an EWS for abrupt dieback events. This behaviour can be expected in the lead up to a vegetation dieback shift because the length and intensity of the dry season has been shown to increase due to drying in the Amazon \citep{9,10}. 

\begin{figure*}[t]
	\includegraphics[width=1\linewidth]{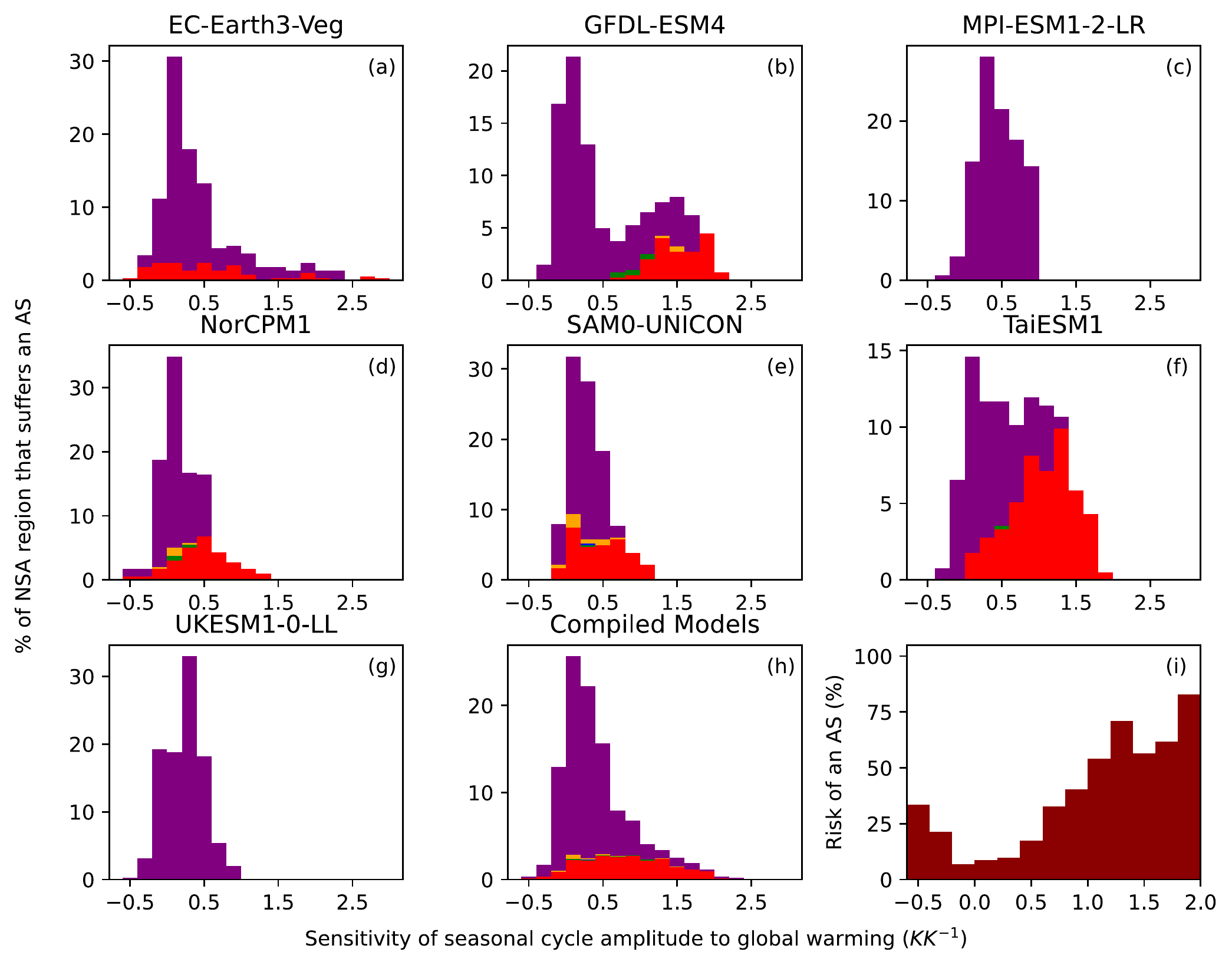}
	\caption{\textbf{Plots demonstrating the efficacy of the temperature seasonal cycle amplitude as an EWS and how it may be used to indicate the risk of an oncoming abrupt dieback shift.} \textbf{(a-g)} Histograms with bins of width 0.2 showing the percentage area of the NSA region that have different sensitivities of the temperature seasonal cycle amplitude to global warming. The bar colors correspond to the type of abrupt shift detected at the grid point with a specific sensitivity, as in Figure 1, where purple bars indicate grid points where no abrupt shifts are detected. \textbf{(h)} The mean of histograms a-g. \textbf{(i)} A bar chart showing how the percentage risk of a grid point, in any of the 7 analysed models, experiencing a dieback shift changes with increasing sensitivity of the seasonal cycle amplitude to global warming}
	\label{Fig_4}
\end{figure*}

Figure~\ref{Fig_4} investigates the robustness of using the temperature seasonal cycle amplitude as an EWS for an impending dieback event. Specifically, we compare the distributions of grid points possessing an abrupt shift and those without for the temperature seasonal cycle amplitude sensitivity to global warming. The sensitivity is calculated up to a doubling of $CO_2$ (most abrupt shifts are detected after doubling of $CO_2$) for all grid points and models (Fig.~\ref{Fig_4}a-g). The grid point are colour coded according to whether an abrupt shift occurs and its type. Four of the five models that contain abrupt shifts within the NSA region (GFDL-ESM4, NorCPM1, SAM0-UNICON, TaiESM1; Fig.~\ref{Fig_4}b,d,e,f) display clear thresholds in the sensitivity, such that above the thresholds only grid points with an abrupt shift feature. EC-Earth3-Veg (Fig.~\ref{Fig_4}a) provides an exception, however due to the high stochasticity of the model some of the detected abrupt shifts are likely to be simply natural variation. Promisingly, MPI-ESM1-2-LR and UKESM1-0-LL (Fig.~\ref{Fig_4}c,g) do not possess grid points with high sensitivities and could therefore offer an explanation for not exhibiting any abrupt shifts in the NSA region.

Taking an ensemble mean of all the models, shows that grid points without an abrupt shift tend to have sensitivities centred around zero, whereas abrupt shift grid points are positively skewed to higher sensitivities (see Fig.~\ref{Fig_4}h). This means that the risk of a grid point having an abrupt shift (defined as the ratio of gird points with an abrupt dieback shift to all grid points for each sensitivity) generally increases for grid points with higher sensitivities to global warming as shown in Fig.~\ref{Fig_4}i. The minimum risk of a grid point experiencing an abrupt shift is for a sensitivity close to 0, where the seasonal cycle amplitude is unaffected by global warming. As the sensitivity increases from 0.5 to 1.0$K/K$ the risk of a grid point containing an abrupt shift increases approximately linearly from 10\% to 60\%. For sensitivities greater than 1.0$K/K$ the risk remains between 60\% and 80\% providing a potential EWS.The risk also increases to 35\% for negative sensitivities, however this is largely from the EC-Earth3-Veg model in which it is not clear how many of the detected shifts are indeed abrupt.

\section{Discussion}
The effects of abrupt shifts observed in the NSA region may be limited if anthropogenic climate change is restricted to below 1.5 degrees, as set out in the aims of the Paris Agreement \citep{23}. Exceeding 1.5°C warming is likely to result in sharp increases in the areas experiencing abrupt shifts. Despite large areas of the Amazon experiencing tipping events with warming, the abrupt shifts observed in Fig.~\ref{Fig_2} may be considered localised events. These are largely balanced out by the increase in vegetation carbon seen elsewhere in the NSA region, likely resulting from CO$_2$ fertilization, where an increase in photosynthesis rate results in an increase in biomass \citep{19,20}. This appears to indicate that large scale regional dieback, as observed in previous generations of models, are not present in CMIP6 and the impacts will be more localised. Thus, despite the CMIP6 models failing to agree on the overall impact of vegetation carbon in the Amazon, abrupt shifts remain a threat to local communities and ecosystems. 

Our analysis shows that typically the sensitivity of the amplitude of the temperature seasonal cycle to global warming is higher for grid points subsequently featuring an abrupt dieback shift, compared with grid points with no abrupt dieback (Fig.~\ref{Fig_4}). This therefore offers the possibility of using this sensitivity as a system specific early-warning signal for future dieback in the Amazon. The increase in risk observed for negative sensitivities also could mean that any changes in moisture and temperature cycles in the Amazon suggest an increased risk of an abrupt shift occurring. 

We find evidence of clustered localised abrupt dieback shifts in over half the CMIP6 models analysed, however, this analysis is limited by the number of CMIP6 models containing dynamic vegetation. Additionally, we use the idealised 1\% per year increase of CO$_2$ run to focus on abrupt dieback shifts caused solely by anthropogenic climate change, though abrupt dieback can also be caused through land use changes such as deforestation.

\section{Conclusions}

Anthropogenic climate change could result in localised tipping events occurring in the Amazon rainforest, as observed in several CMIP6 models. The dieback events detected would have severe consequences for local communities and ecosystems. This study suggests that  7+/-5 $\%$ of the Northern South America region would experience abrupt downward shifts in vegetation carbon per $^{o}$C of global warming above 1.5$^{o}$C.

Further research is needed to assess the risk of tipping events under climate change and to  identify forewarning methods that can be applied to observational data. However, our results indicate that the sensitivity of the amplitude of the temperature seasonal cycle to global warming is a promising potential early warming signal for local Amazon forest dieback. 

\dataavailability{The CMIP6 model output datasets analysed during this study are available online at [https://esgf-node.llnl.gov/search/cmip6/]}

\authorcontribution{IP carried out the data analysis and drafted the paper. PR and PC advised on the stufy. All authors contributed to the submitted paper.} 

\competinginterests{No competing interests} 



\bibliographystyle{copernicus}
\bibliography{Ref.bib}
\end{document}